\documentclass[11pt,draftcls,onecolumn]{IEEEtran}
\ifCLASSINFOpdf
  \usepackage[pdftex]{graphicx}
  \graphicspath{{../pdf/}{../jpeg/}}
  \DeclareGraphicsExtensions{.pdf,.jpeg,.png}
\else
  \usepackage[dvips]{graphicx}
  \graphicspath{{../eps/}}
  \DeclareGraphicsExtensions{.eps}
\fi

\usepackage{graphicx}
\usepackage{epstopdf}

\usepackage{epsfig}
\usepackage{booktabs}

\usepackage{mathtools}

\usepackage{algorithmic}

\usepackage{array}

\usepackage{bm}
\usepackage{mathrsfs}
\usepackage{epsfig}
\usepackage{subfigure}
\usepackage{algorithmic}
\usepackage{mathrsfs}
\usepackage{subfigure}

\usepackage{amsthm}
\usepackage{amssymb}
\usepackage{amsmath}
\usepackage{cite}
\usepackage{url}
\usepackage{xcolor}
\usepackage{cite,graphicx,amsmath,amssymb}
\usepackage{subfigure}
\usepackage{fancyhdr}
\usepackage{mdwmath}
\usepackage{mdwtab}
\usepackage{caption}
\usepackage{amsthm}
\usepackage[square, comma, sort&compress, numbers]{natbib}

\newtheorem{theorem}{Theorem}

\newtheorem{lemma}{Lemma}

\newtheorem{corollary}{Corollary}

\begin{document}

\title{{\color{black}{On the Secrecy of UAV Systems With Linear Trajectory}}
\thanks{Manuscript received**, 2019; revised **, 2019; accepted **, 2019. This research was supported by the NSF of China under Grant 61620106001 and 6180011907, the Open Fund of the Shaanxi Key Laboratory of Information Communication Network and Security under Grant ICNS201807.  The associate editor coordinating the review of this paper and approving it for publication was ***. (Corresponding author: Jianping An.)}
\thanks{G. Pan is with the School of Information and Electronics Engineering, Beijing Institute of Technology, Beijing 100081, China, and he is also with CEMSE Division, King Abdullah University of Science and Technology (KAUST), Thuwal 23955-6900, Saudi Arabia.}
\thanks{H. Lei is with the School of Communication and Information Engineering, Chongqing University of Posts and Telecommunications, Chongqing 400065, China, and he is also with Shannxi Key Laboratory of Information Communication Network and Security, Xi'an University of Posts and Telecommunications, Xi'an, Shaanxi 710121, China.}
\thanks{J. An is with the School of Information and Electronics Engineering, Beijing Institute of Technology, Beijing 100081, China. (e-mail: an@bit.edu.cn)}
\thanks{S. Zhang is with National Key Laboratory of Science and Technology on Aerospace Intelligence Control, Beijing 100854, China, and he is also with Beijing Aerospace Automatic Control Institute, Beijing 100854, China.}
\thanks{M.-S. Alouini is with CEMSE Division, King Abdullah University of Science and Technology (KAUST), Thuwal 23955-6900, Saudi Arabia.}
}

\author{Gaofeng Pan, $Senior Member, IEEE$, Hongjiang Lei, $Member, IEEE$, Jianping An, $Member, IEEE$, Shuo Zhang, and Mohamed-Slim Alouini, $Fellow, IEEE$ }



\maketitle

\begin{abstract}
~{\color{black}{{By observing the fact that moving in a straight line is a common flying behavior of unmanned aerial vehicles (UAVs) in normal applications, e.g., power line inspections, and air patrols along with highway/streets/borders, in this paper we investigate the secrecy outage performance of a UAV system with linear trajectory, where a UAV ($S$) flies in a straight line and transmits its information over the downlink to a legitimate receiver ($D$) on the ground while an eavesdropping UAV ($E$) trying to overhear the information delivery between $S$ and $D$. Meanwhile, some information is delivered to $S$ over the uplink from $D$, such as commanding messages to control $S$'s detecting operations, which can also be eavesdropped by $E$.}}} The locations of $S$, $D$, and $E$ are randomly distributed. We first characterize the statistical characteristics (including cumulative distribution functions and probability density function) of the received signal-to-noise ratio over both downlink and uplink, and then the closed-form analytical expressions for the lower boundary of the secrecy outage probability of both downlink and uplink have also been derived accordingly. Finally, Monte-Carlo simulations are given to testify our proposed analytical models.

\end{abstract}



\section{Introduction}
{\color{black}{Benefiting from the unique characteristics, e.g., fast deployment, easy programmability, reconfiguration, control flexibility, and scalability, unmanned aerial vehicle (UAV) gets more and more popular and are always used for inspection and supervision purposes, such as power line inspection, maritime/harbor/border/highway/street patrol, and police surveillance (like scouting property and locating fugitives) \cite{SHayat,Erdelj,ZengY,Mozaffari,ZZhou,VNNguyen}. In these application scenarios, moving in a straight line is a common flying behavior for UAVs, leading to the linear distribution of the positions of the UAVs, which is different from some other cases that UAVs can locate at any positions in the target zones.}}

{\color{black}{Similar to common territorial wireless systems, information security problems also exist in UAV communication systems, as open wireless channels are employed to deliver information.}} Observing the fact that wireless security can be enhanced in the physical layer, rather than only relying upon generic higher-layer cryptographic mechanisms, physical layer security has been broadly considered as a practicable approach to protecting the data confidentiality from eavesdropping in wireless communication systems \cite{ZXing,YCao,ZXiang,HLeiTCOM2018}, and wireless optical communication systems \cite{GPanTcom2017,GPanCL2017,HLeiPJ2018}. Recently, some researchers pay their attenuations on realizing and enhancing the secure information delivery in common UAV communication systems in term of cooperative transmission, trajectory design, UAV placement, power control, jamming design, and performance modeling \cite{SWKim,YChen,XSun,YCai,ZhangG,SGoel,HLee,LeiIOT,YetWCL,YeGlobecom,JTang}.

As UAV can be flexibly placed in the sky, it is naturally employed to set up or improve the quality of the information delivery over the link between a source and a destination, which suffers deep fading or obstacles, to realize future networks, e.g., sustainable access networks and ultra-dense heterogeneous networks \cite{KYang,JAn}. Then, similar to traditional cooperative systems on the ground, under this case the probability of information leakage always increases, as introducing UAV relay will inevitably increase the opportunity for the eavesdropper to overhear the transmitted information. In \cite{SWKim}, security was studied from multiple UAV control perspective, namely, spatially secure group communication problem is presented and solved to maximize spatial UAV group size while minimizing the communication boundary of the group. To promise the secrecy requirement, it is important to choose the location of the UAV relay, which was studied in Ref. \cite{YChen}. The authors of Ref. \cite{XSun} investigated the secure transmissions of millimeter-wave simultaneous wireless information and power transfer UAV relay networks.

On the other handle, due to the growing cost of infrastructure and the increasing requirements on the performance, UAV's mobility can be exploited to fulfill the requirements arisen day and day. So, the transmission quality between UAV and ground users can be greatly improved via carefully adjusting or calculating UAV's trajectory, while achieving a reasonable balance between the cost and the performance. Ref. \cite{YCai} jointly adjusted UAV trajectories and user scheduling to maximize the minimum worst-case secrecy rate among the users within each period under various constraints, e.g., the maximum UAV speed constraints, UAV return constraints, UAV collision avoidance constraints, etc. In \cite{ZhangG}, the high mobility of UAV was exploited to proactively establish favorable and degraded channels for the legitimate and eavesdropping links, and then to maximize the average secrecy rates of UAV-to-ground and ground-to-UAV transmissions, UAV's trajectory and the transmit power of the legitimate transmitter were jointly optimized.

Moreover, jamming has been widely regarded as a realistic method to improve the secrecy performance of wireless communication systems. Then, the jamming method has also been introduced into UAV communication systems. In \cite{SGoel}, the source was designed to transmit artificial noise signals, in addition to information signals, to confuse this eavesdropper. In \cite{HLee}, a cooperative UAV was considered to transmit the jamming signal. Furthermore, the authors maximized the minimum secrecy rate among the ground users by jointly optimizing the trajectory and the transmit power of the UAVs as well as the user scheduling.

Since performance modeling is an efficient way to study and understand the performance of communication systems, researchers set up mathematic models to investigate the secrecy performance of UAV communication systems. The influence of the randomness of the location of UAV has been studied for UAV-to-UAV and UAV-to-ground links while suffering multiple eavesdropping UAVs in \cite{YetWCL,LeiIOT,YeGlobecom}. The analytical expression was derived for the secure connection probability of the legitimate ground link in the presence of non-colluding UAV eavesdroppers \cite{JTang}.

{\color{black}{Therefore, it is clear that no works have been presented to study the secure information delivery in UAV systems with linear trajectory, though there are some works proposed for various kinds of UAV communication systems. To fulfill this blank and understand how system factors affect the secrecy performance, it is necessary to investigate the secure information transmission in UAV systems with linear trajectory. In this work, a UAV system with a linear trajectory is considered, where a UAV ($S$) is adopted to perform inspection/supervision. Especially, $S$ moves in a straight line (e.g., power line inspections, and air patrols along with highway/streets/borders), and sends back inspection data to a ground receiver ($G$).}} Simultaneously, there is an eavesdropping UAV ($E$) trying to overhear the information delivery between $S$ and $G$.

Similar to \cite{PanGlobecom17,PanTGCN19}, the impacts of the randomness of $S$ and $E$ in 3-dimensional space, and that of $G$ on the ground on the secrecy outage performance of the considered UAV system with linear trajectory are investigated by employing stochastic geometry theory.

The main contributions of this paper are summarized as follows:

1) We characterize the statistical characteristics (including cumulative distribution functions (CDF) and probability density function (PDF)) of the received signal-to-noise ratio (SNR) at $E$, $G$ over the downlink, and at $S$ over the uplink, respectively;

2) We derive the closed-form analytical expressions for the lower boundary of the secrecy outage probability (SOP) of uplink and downlink, respectively;

3) We systematically study and summarize the impacts of the radius of the coverage space of $S$ and $G$, the transmit SNR at $S$ and $G$, and the height of UAV on the secrecy performance of the considered system.

The rest of this paper is organized as follows. In Section II, the considered power-line inspection UAV system is described. In Section III and IV, the secrecy outage analysis is conducted for both the uplink and the downlink, respectively. Also, closed-form analytical expressions for the lower boundary of the SOP of both uplink and downlink are derived. In Section V, numerical results for the secrecy outage are presented and discussed. Finally, we conclude the paper with some remarks in Section VI.

\section{System Model}
\begin{figure}[!htb]
\centering
\includegraphics[width= 3.5in,angle=0]{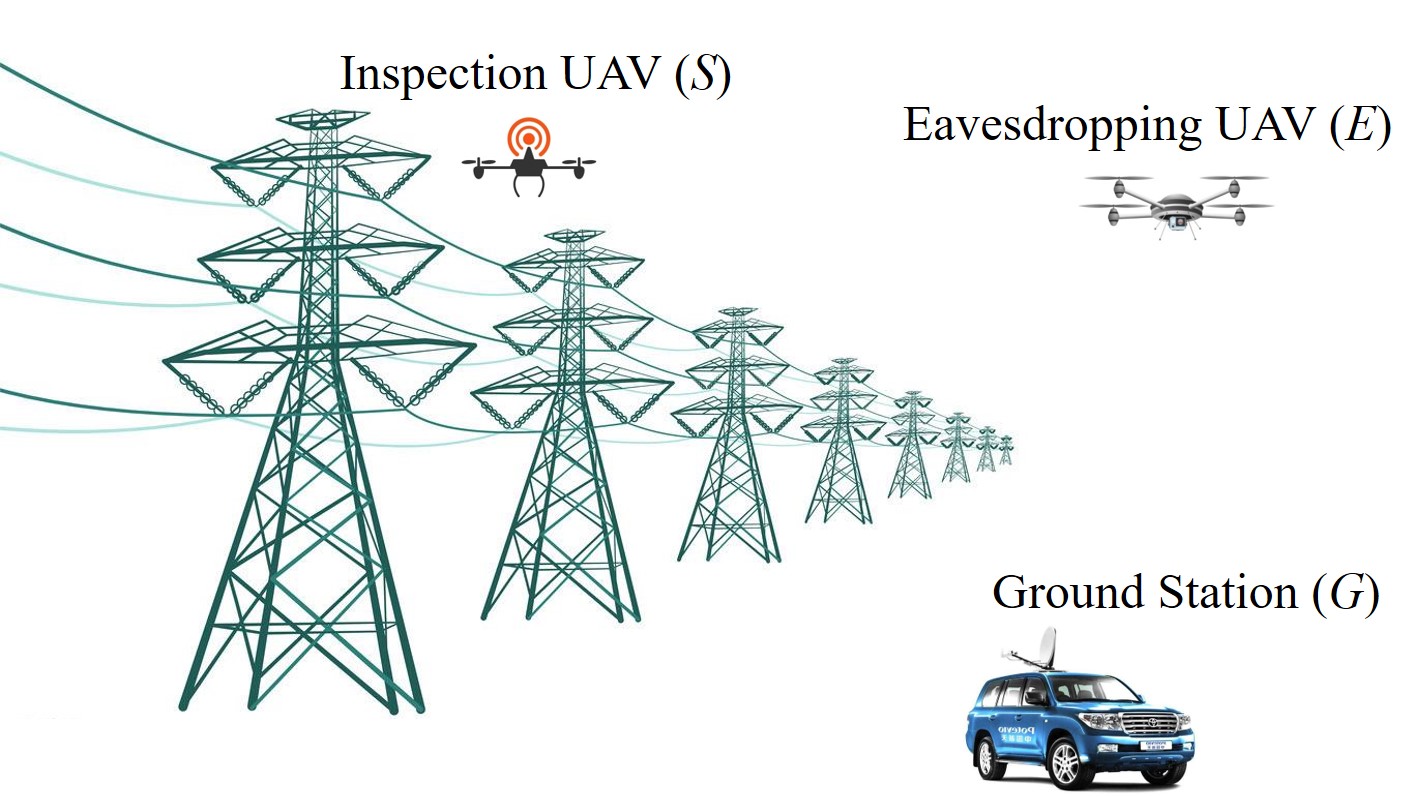}
\caption{Automatic power-line inspection UAV system.}
\label{fig_1}
\end{figure}

{\color{black}{In this paper, a UAV system with linear trajectory is considered, in which $S$ is adopted to perform inspection/supervision. Specially, $S$ moves in a straight line (e.g., power line inspections, and air patrols along with highway/streets/borders), and sends back inspection data to $G$. Simultaneously, $E$ tries to overhear the information transmissions between $S$ and $G$. In other words, $E$ can eavesdrop the information delivery over both the uplink link from $G$ to $S$ and the downlink from $S$ to $G$. To facilitate the illustration, we take power line inspection as an example of the considered system, shown in Fig. \ref{fig_1}.}} Furthermore, an omnidirectional transmission antenna is assumed to be employed at both $S$ and $G$. In this work, we also assume that all links suffer independent Rayleigh fading to reflect the effects of small-scale fading, namely, the channel gain ${h_{jk}} \sim {\cal C}{\cal N}\left( {0,{g_{jk}}} \right)$, where ${g_{jk}} = E\left( {{{\left| {{h_{jk}}} \right|}^2}} \right)$ and $j,k \in \{ G,S,E\} $\footnote{{\color{black}{In this work, non-line-of-sight (NLOS) propagation scenarios (e.g., UAVs are employed to patrol and monitor the urban streets) are considered to facilitate presenting the technical road-map on analyzing the SOP of the considered system, which can serve as a useful reference to study the performance of similar systems, no matter in NLOS or line-of-sight propagation scenarios.}}}.

In this work, it is assumed that the trajectory of $S$ is in a straight line for tractability purposes. This assumption is reasonable, as $S$ normally flies along with highway/streets/borders to inspect/monitor and gather the state information. Furthermore, we also assume that $E$ is uniformly distributed in the airspace to eavesdrop the information delivery between $S$ and $G$\footnote{{\color{black}{In this work, passive eavesdropping is considered, which can realize the optimal eavesdropping from $E$'s side. If the operating space of the legitimate users is open, $E$ can share this space with legitimate users and pretend to be the legal user, and then it can approach the legitimate users with no doubts. Therefore, here we do not consider the minimum distance between $E$ and legitimate users for simplification.}}}.

\begin{figure}[!t]
\centering
\includegraphics[width= 3.5in,angle=0]{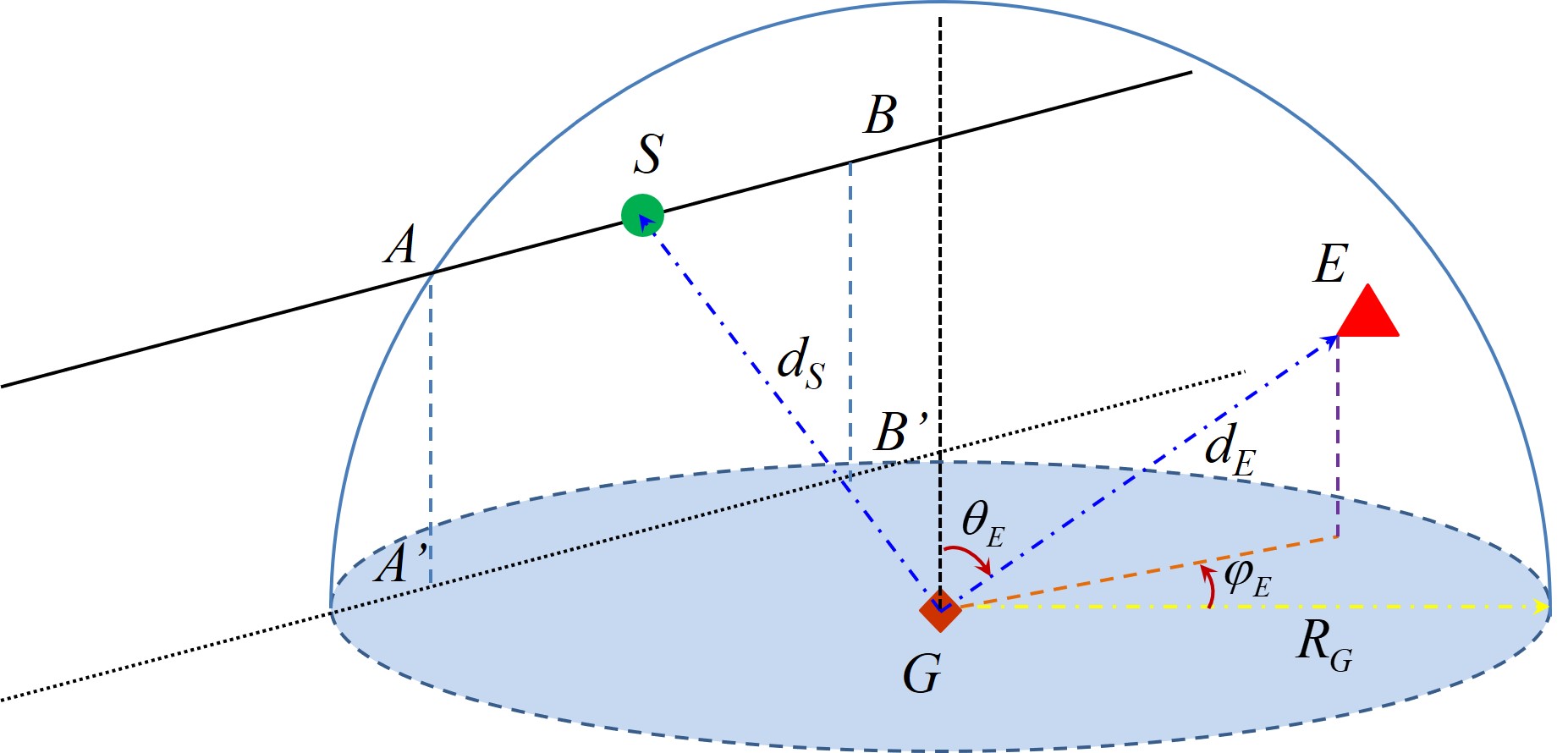}
\caption{Uplink model.}
\label{fig_2}
\end{figure}

\begin{figure}[!t]
\centering
\includegraphics[width= 2.5in,angle=0]{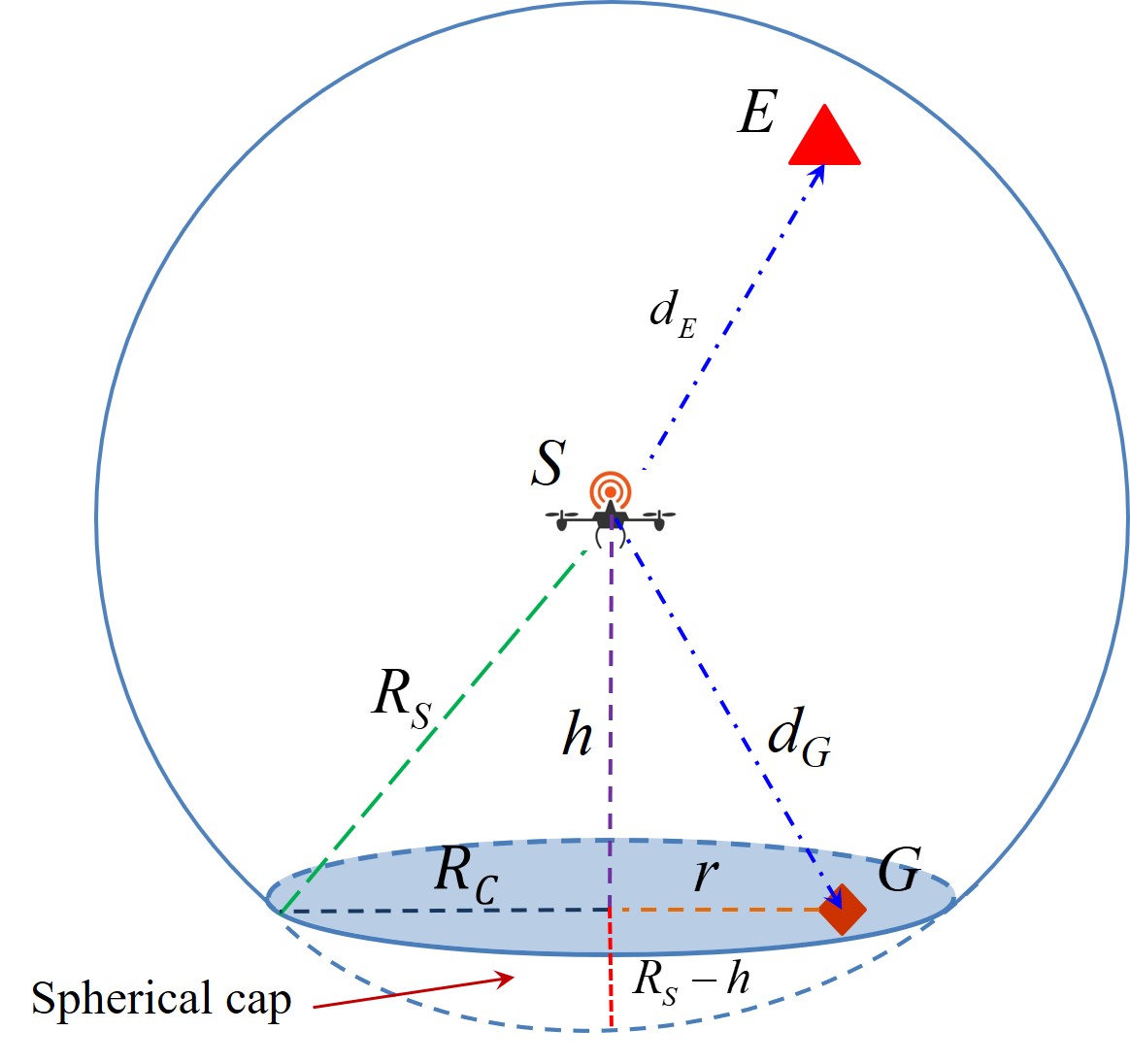}
\caption{Downlink model.}
\label{fig_3}
\end{figure}

As depicted in Fig. \ref{fig_2}, during the uplink transmission stage, the coverage space of $G$ is a hemisphere with radius $R_G$, the center of which is $G$\footnote{{\color{black}{In practical, UAVs suffer their minimum flying heights ranging from meters or more, which depends on the designs and application purposes. Similar to the analysis on traditional wireless systems, in order to facilitate theoretical analysis, in this work the minimum operating heights of the UAVs and the height of the transceiver antennas at $G$ are ignored.}}}. As presented in Fig. \ref{fig_3}, in the downlink delivery stage, the coverage space of $S$ is a spherical cap with height $R_S + h$ (where ${R_C} = \sqrt {R_S^2 - h_{}^2} $) and the radius of the base $R_C$ (where $0 \le h \le {R_S}$). Moreover, it is easy to obtain the volumes of the coverage spaces of $G$ and $S$ as ${V_{G}} = \frac{2}{3}\pi R_G^3$ and ${V_{{S_1}}} = \frac{\pi }{3}\left( {4R_S^3 - 3{R_S}h^2 + h^3} \right)$, respectively.

In the following two sections, secrecy outage analysis will be presented for the information delivery over uplink and downlink, respectively.

\section{Secrecy Outage Analysis for The Uplink}
As depicted in Fig. \ref{fig_2}, it is assumed that $S$ is uniformly distributed in a straight-line segment $AB$, the length of which $l$ ($0 < l \le 2R_G^{}$) is determined by the juncture points of power-line and the coverage space of $G$. $E$ is uniformly distributed in the hemisphere with radius $R_G$ and center $G$.

\subsection{Signal model}
The received signals at $S$ and $E$ can be respectively written as
\begin{align}
{y_S} = \sqrt {{P_G}d_S^{ - n}} {h_{GS}}{s_G} + {z_S}
\end{align}and
\begin{align}
{y_E} = \sqrt {{P_G}d_E^{ - n}} {h_{GE}}{s_G} + {z_{_E}},
\end{align}
where ${P_G}$ is the transmit power at $G$, $s_G$ denotes the transmitted symbols from $G$, $d_S$ is the distance between $G$ and $S$, $d_E$ is the distance between $G$ and $E$, $n$ is the path-loss exponent, $z_S$ and $z_E$ denote the independent complex Gaussian noise at $S$ and $E$, respectively. In this work, to simplify the analysis, we assume that $z_S$ and $z_E$ are with zero means and a same variances, ${N_0}$.

Therefore, the received SNR at $S$ and $E$ can be further given as
\begin{align}\label{SNRS}
{\gamma _S} = \frac{{{P_G}{{\left| {{h_{GS}}} \right|}^2}}}{{d_S^n{N_0}}} = {\lambda _G}\frac{{{{\left| {{h_{GS}}} \right|}^2}}}{{d_S^n}}
\end{align}and
\begin{align}\label{SNRE}
{\gamma _E} = \frac{{{P_G}{{\left| {{h_{GE}}} \right|}^2}}}{{d_E^n{N_0}}} = {\lambda _G}\frac{{{{\left| {{h_{GE}}} \right|}^2}}}{{d_E^n}},
\end{align}
respectively, where ${\lambda _G} = \frac{{{P_G}}}{{{N_0}}}$.

As in this work all links suffer independent Rayleigh fading, we can obtain the PDF of ${\left| {{h_{jk}}} \right|^2}$ ($j,k \in \{ G,S,E\} $) as
\begin{align}\label{pdfrayleigh}
{f_{{{\left| {{h_{jk}}} \right|}^2}}}\left( x \right) = \frac{1}{{{g_{jk}}}}\exp \left( { - \frac{x}{{{g_{jk}}}}} \right),
\end{align}
where ${g_{jk}}$ is the mean value of the power gain, ${{\left| {{h_{jk}}} \right|}^2}$.

\begin{figure}[!t]
\centering
\includegraphics[width= 1.5in,angle=0]{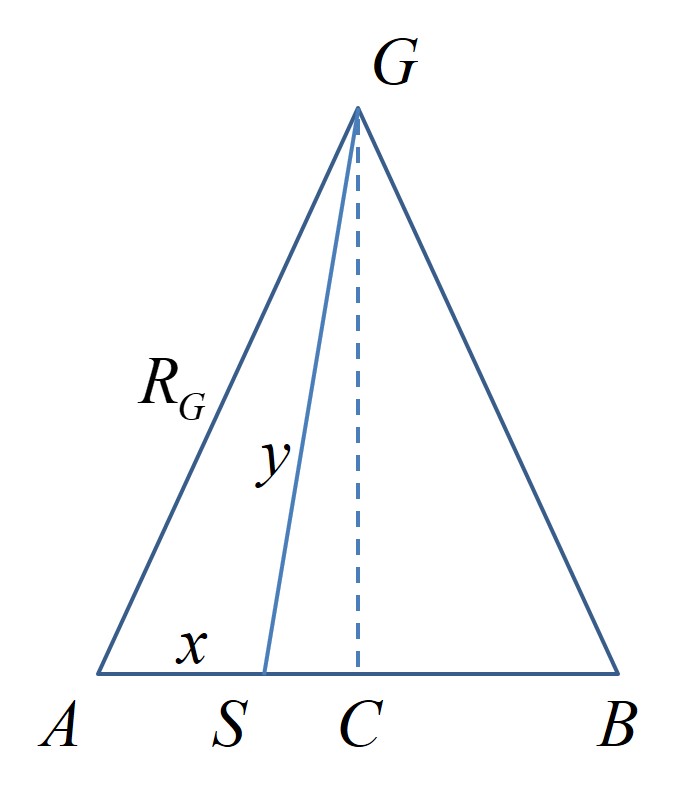}
\caption{An isosceles triangle.}
\label{fig_4}
\end{figure}
\subsection{The Derivation of The CDF of $\gamma_S$}
In order to facilitate the following analysis, we first give a useful theorem to characterize the statistical characteristics of $d_S$ as follows:

\begin{theorem}
{As shown in Fig. \ref{fig_4}, given an isosceles triangle $\triangle GAB$, in which $GA = GB = b$, $AB = l$, $S$ is uniformly distributed over $AB$, $C$ is the midpoint of $AB$, $AS = x$ ($0 \le x \le l$) and $GS = y$ ($c = \sqrt {{b^2} - \frac{{{l^2}}}{4}}  \le y \le b$). Thus, the PDF of $y$ can be written as}
\begin{align}\label{theorem1}
{f_Y}\left( y \right) = \left\{ {\begin{array}{*{20}{l}}
{\frac{{4y}}{{l\sqrt {4{y^2} + {l^2} - 4{b^2}} }}}&{{\rm{if}}\sqrt {{b^2} - \frac{{{l^2}}}{4}}  \le y \le b}\\
0&{{\rm{else}}}
\end{array}} \right..
\end{align}

\begin{proof}
\emph{Please refer to Appendix A. }
\end{proof}
\end{theorem}

\begin{corollary}
{As shown in Fig. \ref{fig_2}, $S$ is uniformly distributed in $AB$ and $G$ is at the centra of the hemisphere. Then, the the PDF of the distance between $S$ and $G$, $d_S$, can be given as}
\begin{align}\label{pdfdS}
{f_{{d_S}}}\left( x \right) = \left\{ {\begin{array}{*{20}{l}}
{\frac{{4x}}{{l\sqrt {4{x^2} + {l^2} - 4{b^2}}, }}}&{{\rm{if}}\sqrt {b^2 - \frac{{{l^2}}}{4}}  \le x \le b}\\
0,&{{\rm{else}}}
\end{array}} \right..
\end{align}
\begin{proof}
\emph{Eq. \eqref{pdfdS} can be easily achieved by applying \textbf{\emph{Theorem 1}}.}
\end{proof}
\end{corollary}

\begin{corollary}
{Considering the randomness of the position of $S$, the PDF of $d_S^n/{\lambda _G}$ can be presented as}
\begin{equation}
{f_{\frac{{d_S^n}}{{{\lambda _G}}}}}\left( x \right) = \left\{ {\begin{array}{*{20}{c}}
{\frac{{2{\lambda _G}}}{{nl}}\frac{{{{\left( {{\lambda _G}x} \right)}^{\frac{2}{n} - 1}}}}{{\sqrt {{{\left( {{\lambda _G}x} \right)}^{\frac{2}{n}}} - c} }},}&{\frac{1}{{{\lambda _G}}}{c^{\frac{n}{2}}} < x < \frac{{{b^n}}}{{{\lambda _G}}}}\\
0,&{{\rm{else}}}
\end{array}} \right..
\label{pdfdsn}
\end{equation}

\begin{proof}
\emph{Using \textbf{\emph{Corollary 1}}, the CDF of $d_S$ can be given as
\begin{align}
{F_{{d_S}}}\left( x \right) &= \int\limits_0^x {{f_{{d_S}}}\left( y \right)} dy\notag\\
& = \left\{ {\begin{array}{*{20}{l}}
0&{x < \sqrt c }\\
{\frac{{2\sqrt {{x^2} - c} }}{l}}&{\sqrt c  < x < b}\\
1&{y > b}
\end{array}} \right.,
\end{align}where $c = {b^2} - \frac{{{l^2}}}{4}$.}

\emph{So, one can have the CDF of $d_S^n/{\lambda _G}$ as}
\begin{align}\label{CDFdsn}
{F_{\frac{{d_S^n}}{{{\lambda _G}}}}}\left( x \right) &= \left\{ {\frac{{d_S^n}}{{{\lambda _G}}} \le x} \right\}\notag\\
& = \left\{ {{d_S} \le \sqrt[n]{{{\lambda _G}x}}} \right\}\notag\\
& = \left\{ {\begin{array}{*{20}{l}}
0&{x < \frac{1}{{{\lambda _G}}}{c^{\frac{n}{2}}}}\\
{\frac{{2\sqrt {{{\left( {x{\lambda _G}} \right)}^{\frac{2}{n}}} - c} }}{l}}&{\frac{1}{{{\lambda _G}}}{c^{\frac{n}{2}}} < x < \frac{{{b^n}}}{{{\lambda _G}}}}\\
1&{x > \frac{{{b^n}}}{{{\lambda _G}}}}
\end{array}} \right..
\end{align}

\emph{Therefore, the PDF of $d_S^n/{\lambda _G}$ can be obtained, as it is the derivation of Eq. \eqref{CDFdsn}.}

\emph{Then, the proof is completed.}
\end{proof}
\end{corollary}

Observing Eq. \eqref{SNRS}, and using Eqs. \eqref{pdfrayleigh} and \eqref{pdfdsn}, the CDF of $\gamma_S$ can be calculated as
\begin{equation}
\begin{aligned}
{F_{{\gamma _S}}}\left( \gamma  \right) &= \Pr \left\{ {{\lambda _G}\frac{{{{\left| {{h_{GS}}} \right|}^2}}}{{d_S^n}} < \gamma } \right\}\\
& = \Pr \left\{ {{{\left| {{h_{GS}}} \right|}^2} < \frac{{d_S^n}}{{{\lambda _G}}}\gamma } \right\}\\
& = \int_0^\infty  {{F_{{{\left| {{h_{GS}}} \right|}^2}}}\left( {t\gamma } \right){f_{\frac{{d_S^n}}{{{\lambda _G}}}}}\left( t \right)dt} \\
& = 1 - \frac{{2{\lambda _G}}}{{nl}}\int_{\frac{{{c^{\frac{n}{2}}}}}{{{\lambda _G}}}}^{\frac{{{b^n}}}{{{\lambda _G}}}} {\exp \left( { - \frac{{t\gamma }}{{{g_{GS}}}}} \right)\frac{{{{\left( {{\lambda _G}t} \right)}^{\frac{2}{n} - 1}}}}{{\sqrt {{{\left( {{\lambda _G}t} \right)}^{\frac{2}{n}}} - c} }}dt}.
\label{H232}
\end{aligned}
\end{equation}

For analytical tractability, in this work, we only consider cases that the path loss factor is $n = 2$, which is suitable for the case of infinite space and match the open air space scenarios considered in this work, as suggested in \cite{Rappaport}.

Then, we can further calculate the CDF of $\gamma_S$ as
\begin{equation}
\begin{aligned}
{F_{{\gamma _S}}}\left( \gamma  \right) &= 1 - \frac{{{\lambda _G}}}{l}\int_{\frac{c}{{{\lambda _G}}}}^{\frac{{{b^2}}}{{{\lambda _G}}}} {\exp \left( { - \frac{{t\gamma }}{{{g_{GS}}}}} \right)\frac{1}{{\sqrt {{\lambda _G}t - c} }}dt} \\
& = 1 - \frac{{\sqrt {{\lambda _G}} }}{l}\int_{\frac{c}{{{\lambda _G}}}}^{\frac{{{b^2}}}{{{\lambda _G}}}} {\exp \left( { - \frac{{t\gamma }}{{{g_{GS}}}}} \right)\frac{1}{{\sqrt {t - \frac{c}{{{\lambda _G}}}} }}dt} \\
& = 1 - A_S{\gamma ^{ - \frac{1}{2}}}\exp \left( { - B_S\gamma } \right){\mathop{\rm erf}\nolimits} \left( {C_S{\gamma ^{\frac{1}{2}}}} \right),
\label{CDFgammaS}
\end{aligned}
\end{equation}
where $A_S = \frac{{2\sqrt {\pi {g_{GS}}{\lambda _G}} }}{l}$,
$B_S = \frac{c}{{{g_{GS}}{\lambda _G}}}$, and
$C_S = \sqrt {\frac{{{b^2} - c}}{{{g_{GS}}{\lambda _G}}}} $.

\subsection{The Derivation of The PDF of $\gamma_E$}
\begin{lemma}\label{lemma1}
{As $E$ is uniformly distributed in the coverage space of $S$, the PDF of $d_E$ can be written as}
\begin{align}\label{fdE}
{f_{{d_E}}}\left( x \right) = \left\{ \begin{array}{l}
\frac{{3{x^2}}}{{{ R_G ^3}}},\; {\rm{if}} \;0 \le x \le R_G  \\
0,\;\;\;\; {\rm{else}}
\end{array} \right..
\end{align}
\begin{proof}
\emph{The PDF of the distance between $E$ and $S$ can be easily calculated as $f_D(x)  = \frac{3} {2\pi{ R_G ^3}}$.
Therefore, we can have the CDF of $d_E$ as}
\begin{align}\label{FdE}\notag
{F_{{d_E}}}\left( x \right)& = \int\limits_0^x {\int\limits_0^{\pi/2} {\int\limits_0^{2\pi } {\frac{3}{{2\pi {R_G^3}}}\sin {\phi _i}d_i^2\mathrm{d}{\theta _i}\mathrm{d}{\phi _i}\mathrm{d}\left( {{d_i}} \right)} } } \\
 &= \left\{ \begin{array}{l}
0,\;\;\;\;x < R_G \\
\frac{{{x^3}}}{{{ R_G ^3}}},\;0 \le x \le R_G \\
1,\;\;\;\;x > R_G
\end{array} \right..
\end{align}

\emph{Thus, one can easily achieve the PDF of $d _E$ as the derivative of Eq. \eqref{FdE}.}

\emph{Then, the proof is completed.}
\end{proof}
\end{lemma}

So, it is easy to have the CDF of $d_E^n/{\lambda _G}$ as
\begin{align}
{F_{\frac{{d_E^n}}{{{\lambda _G}}}}}\left( x \right) &= \left\{ {\frac{{d_E^n}}{{{\lambda _G}}} \le x} \right\}\notag\\
& = \left\{ {{d_E} \le \sqrt[n]{{{\lambda _G}x}}} \right\}\notag\\
& = \left\{ {\begin{array}{*{20}{l}}
{0,}&{x \le 0}\\
{\frac{{{\lambda _G}^{\frac{3}{n}}}}{{R_G^3}}{x^{\frac{3}{n}}}}&{0 < x \le \frac{{{R_G}^n}}{{{\lambda _G}}}}\\
1&{x \ge \frac{{{R_G}^n}}{{{\lambda _G}}}}
\end{array}} \right..
\end{align}

Accordingly, the PDF of $d_E^n/{\lambda _G}$ can be derived as
\begin{align}\label{pdfdEn}
{f_{\frac{{d_E^n}}{{{\lambda _G}}}}}\left( x \right) = \left\{ {\begin{array}{*{20}{l}}
{\frac{{3{\lambda _G}^{\frac{3}{n}}}}{{nR_G^3}}{x^{\frac{3}{n} - 1}}}&{{\rm{if }}~0 < x \le \frac{{{R_G}^n}}{{{\lambda _G}}}}\\
0&{{\rm{else}}}
\end{array}} \right..
\end{align}

Observing Eq. \eqref{SNRE} and using \cite[Eq. (3.351.1)]{Gradshteyn} and Eqs. \eqref{pdfrayleigh} and \eqref{pdfdEn}, the PDF of $\gamma_{E}$ can be calculated as
\begin{align}\label{PDFgammaE}
{f_{{\gamma _E}}}\left( x \right) &= \int\limits_0^\infty  {y{f_{{{\left| {{h_{GE}}} \right|}^2}}}\left( {yx} \right)} {f_{\frac{{d_E^n}}{{{\lambda _G}}}}}\left( y \right)dy\notag\\
&= \frac{{3{\lambda _G}^{\frac{3}{n}}}}{{nR_G^3{g_{GE}}}}\int\limits_0^{\frac{{{R_G}^n}}{{{\lambda _G}}}} y \exp \left( { - \frac{{yx}}{{{g_{GE}}}}} \right){y^{\frac{3}{n} - 1}}dy\notag\\
&=\eta{x^{ - \frac{3}{n} - 1}}\Upsilon \left( {1 + \frac{3}{n},\frac{{{R_G}^n}}{{{\lambda _G}{g_{GE}}}}x} \right),
\end{align}
where $\eta = \frac{{3{\lambda _G}^{\frac{3}{n}}{g_{GE}}^{\frac{3}{n}}}}{{nR_G^3}}$ and
$\Upsilon \left( {\alpha ,x} \right) = \int_0^x {{e^{ - t}}{t^{\alpha  - 1}}dt}$ is the lower incomplete Gamma function, as defined by \cite[(8.350.1)]{Gradshteyn}.

\subsection{The Calculation of SOP}
Therefore, the instantaneous secrecy capacity of the uplink transmission can be presented as
\begin{align}
{C_{s,{\rm{up}}}}\left( {{\gamma _S},{\gamma _E}} \right) = \max \left\{ {{{\log }_2}\left( {1 + {\gamma _S}} \right) - {{\log }_2}\left( {1 + {\gamma _E}} \right)}, 0 \right\}.
\end{align}

In this work, SOP is defined as the probability that the secrecy capacity is less than a outage threshold $R_s$, which can be written as
\begin{equation}
\begin{aligned}
{P_{out,{\rm{up}}}} &= \Pr \left\{ {{C_{s,{\rm{up}}}}\left( {{\gamma _S},{\gamma _E}} \right) \le {R_s}} \right\}\\
& = \Pr \left\{ {{{\log }_2}\left( {1 + {\gamma _S}} \right) - {{\log }_2}\left( {1 + {\gamma _E}} \right) \le {R_s}} \right\}\\
& = \Pr \left\{ {{\gamma _S} \le \Theta {\gamma _E} + \Theta  - 1} \right\},
\label{POUT0}
\end{aligned}
\end{equation}
where $\Theta  = {2^{{R_s}}}$.

It must be noted that obtaining a closed-form result for Eq. (\ref{POUT0}) is almost impossible and/or too complex. Therefore in the subsequent section, the lower bound of the SOP is considered\footnote{{\color{black}{In this work, the lower bound of SOP is considered by exploiting the fact that, compared with $\Theta {\gamma _E} + \Theta  - 1$, a lower outage threshold, $\Theta {\gamma _E}$, represents a more rigorous outage request for ${\gamma _S}$ and is more useful from the engineering perspective.}}}, which has been utilized in many works \cite{Lei2018PJ, Lei2019TVT}, as
{\color{black}{\begin{equation}
\begin{aligned}
P_{out,{\rm{up}}}^{\text{L}} &= \Pr \left\{ {{\gamma _{S}} \leqslant \Theta {\gamma _{E}}} \right\}\\
& = \int_0^\infty  {{F_{{\gamma _{S}}}}\left( {\Theta {\gamma _{E}} } \right)} {f_{{\gamma _{E}}}}\left( {{\gamma _{E}}} \right)d{\gamma _{E}}.
\label{PoutL}
\end{aligned}
\end{equation}}}

Then, using Eqs. \eqref{CDFgammaS} and \eqref{PDFgammaE}, $\Upsilon \left( {a,z} \right) = G_{1,2}^{1,1}\left[ {z\left| {_{a,0}^1} \right.} \right]$,
${\rm{erf}}\left( x \right) = {\pi ^{ - 0.5}}G_{1,2}^{1,1}\left[ {{x^2}\left| {_{0.5,0}^1} \right.} \right]$,
$\exp \left( { - x} \right) = G_{0,1}^{1,0}\left[ {x\left| {_0^ - } \right.} \right]$,
and utilizing the integral equation presented in \cite[(20)]{LeiH2016TVT}, we can calculate SOP as
\begin{equation}
\begin{aligned}
P_{out,{\rm{up}}}^{\rm{L}} &= \int_0^\infty  {{F_{{\gamma _S}}}\left( {\Theta x} \right){f_{{\gamma _E}}}\left( x \right)dx} \\
& = 1 - \frac{{A_S\eta }}{{\sqrt \Theta  }}\int_0^\infty  {{x^{ - 3}}\exp \left( { - B_S\Theta x} \right){\rm{erf}}\left( {C_S\sqrt \Theta  {x^{\frac{1}{2}}}} \right)\Upsilon \left( {\frac{5}{2},\frac{{R_G^2}}{{{\lambda _G}{g_{GE}}}}x} \right)dx} \\
& = 1 - \frac{{A_S\eta }}{{\sqrt {\Theta \pi } }}\int_0^\infty  {{x^{ - 3}}G_{0,1}^{1,0}\left[ {B_S\Theta x\left| {_0^ - } \right.} \right]G_{1,2}^{1,1}\left[ {{C_S^2}\Theta x\left| {_{0.5,0}^1} \right.} \right]G_{1,2}^{1,1}\left[ {\frac{{R_G^2}}{{{\lambda _G}{g_{GE}}}}x\left| {_{2.5,0}^1} \right.} \right]dx} \\
& = 1 - \frac{{A_S\eta {{\left( {{C^2}\Theta } \right)}^3}}}{{\sqrt {\Theta \pi } }}\int_0^\infty  {G_{0,1}^{1,0}\left[ {B_S\Theta x\left| {_0^ - } \right.} \right]G_{1,2}^{1,1}\left[ {{C_S^2}\Theta x\left| {_{ - 2.5, - 3}^{ - 2}} \right.} \right]G_{1,2}^{1,1}\left[ {\frac{{R_G^2}}{{{\lambda _G}{g_{GE}}}}x\left| {_{2.5,0}^1} \right.} \right]dx} \\
& = 1 - \frac{{A_S\eta {C_S^6}{\Theta ^2}}}{{B_S\sqrt {\Theta \pi } }}G_{0,1:1,2:1,2}^{1,0:1,1:1,1}\left( {_ - ^1\left| {_{ - 2.5, - 3}^{ - 2}\left| {_{2.5,0}^1\left| {\frac{{{C_S^2}}}{B_S},\frac{{R_G^2}}{{B_S\Theta {\lambda _G}{g_{GE}}}}} \right.} \right.} \right.} \right)
\label{OPup}
\end{aligned}
\end{equation}
where $G_{p,q}^{m,n}\left[ {x\left| {_{{b_1}, \cdots ,{b_q}}^{{a_1}, \cdots ,{a_p}}} \right.} \right]$ $ = \frac{1}{{2\pi i}}\int {\frac{{\prod\limits_{j = 1}^m {\Gamma \left( {{b_j} - s} \right)} \prod\limits_{j = 1}^n {\Gamma \left( {1 - {a_j} + s} \right)} }}{{\prod\limits_{j = m + 1}^q {\Gamma \left( {1 - {b_j} + s} \right)} \prod\limits_{j = n + 1}^p {\Gamma \left( {{a_j} - s} \right)} }}} {x^s}ds$ is the Meijer's $G$-function, as defined by \cite[Eq. (9.301)]{Gradshteyn}, and
$G_{{p_1},{q_1}:{p_2},{q_2}:{p_3},{q_3}}^{{m_1},{n_1}:{m_2},{n_2}:{m_3},{n_3}}\left[  \cdot  \right]$ is the EGBMGF function, which can be easily realized by utilizing MATHEMATICA${\textregistered}$ (see Table 1 in \cite{Ansari2011TCOM}).


\section{Secrecy Outage Analysis for Downlink}
As shown in Fig. \ref{fig_3}, it is assumed that $G$ is uniformly distributed in the base of the cap with height ($R_S + h$), which is a circle with radius ${R_C} = \sqrt {R_S^2 - h^2} $. In the following, we name the spherical cap shown in Fig. \ref{fig_3} as $S_1$ and its volume as ${V_{{S_1}}} = \frac{\pi }{3}\left( {4R_S^3 - 3{R_S}h^2 + h^3} \right)$.

Therefore, similar to last section, we can obtain the SOP for the downlink transmission as
\begin{equation}
P_{out,{\rm{dn}}}^{\rm{L}} = \int_0^\infty  {{F_{{\gamma _G}}}\left( {\Theta x} \right){f_{{\gamma _E}}}\left( x \right)dx},
\label{H232}
\end{equation}
where ${\gamma _G} = \frac{{{P_S}{{\left| {{h_{SG}}} \right|}^2}}}{{d_G^n{N_0}}} = {\lambda _S}\frac{{{{\left| {{h_{SG}}} \right|}^2}}}{{d_G^n}}$ is the received SNR at $G$, $P_S$ is the transmit power at $S$, ${\lambda _S} = \frac{{{P_S}}}{{{N_0}}}$, ${\gamma _E} = \frac{{{P_S}{{\left| {{h_{SE}}} \right|}^2}}}{{d_E^n{N_0}}} = {\lambda _S}\frac{{{{\left| {{h_{SE}}} \right|}^2}}}{{d_E^n}}$ is the received SNR at $E$.


As depicted in Fig. \ref{fig_3} and considering the randomness of the positions of $G$ and $E$, one can have
\begin{align}\label{Pdown}
P_{out,{\rm{dn}}}^{\rm{L}} &= \int\limits_{{V_{{S_1}}}} {{\rm{Pr\{ }}{\gamma _G} \le \Theta{\gamma _E}{\rm{\} }}} d{V_{{S_1}}}\notag\\
& = \int\limits_{{V_{Sp}}} {{\rm{Pr\{ }}{\gamma _G} \le \Theta{\gamma _E}{\rm{\} }}} d{V_{{S_1}}}\notag\\
&~~ - \frac{{{V_{{S_2}}}}}{{{V_{{S_p}}}}}\int\limits_{{V_{{S_2}}}} {{\rm{Pr\{ }}{\gamma _G} \le \Theta{\gamma _E}{\rm{\} }}} d{V_{{S_2}}}\notag\\
& = {I_{Sp}} - \frac{{{h^2}\left( {3{R_S} - h} \right)}}{{4{R_S}^3}}{I_{{S_2}}},
\end{align}
where ${I_{Sp}} = \int\limits_{{V_{Sp}}} {{\rm{Pr\{ }}{\gamma _G} \le \Theta{\gamma _E}{\rm{\} }}} d{V_{{S_1}}}$, ${I_{{S_2}}} = \int\limits_{{V_{{S_2}}}} {{\rm{Pr\{ }}{\gamma _G} \le \Theta{\gamma _E}{\rm{\} }}} d{V_{{S_2}}}$, $S_p$ is the sphere with radius $R_S$ and volume as ${V_{Sp}} = \frac{4}{3}\pi {R_S}^3$, the spherical cap ($S_2$) with the height ($R_S - h$) and volume ${V_{{S_2}}} = \pi {{h}^2}\left( {{R_S} - \frac{{h}}{3}} \right)$.

In the following two subsections, we will calculate ${I_{Sp}}$ and ${I_{S_2}}$, respectively.

\subsection{The Derivation of ${I_{Sp}}$}

\begin{lemma}\label{lemma2}
{When $E$ is uniformly distributed in the ball with the centra, $S$, the PDF of $d_E^n/{\lambda _S}$ can presented as}
\begin{align}\label{pdfdEn2}
{f_{\frac{{d_E^n}}{{{\lambda _S}}}}}\left( x \right) = \left\{ {\begin{array}{*{20}{l}}
{\frac{{3{\lambda _S}^{\frac{3}{n}}}}{{nR_S^3}}{x^{\frac{3}{n} - 1}}}&{{\rm{if ~}}0 \le x \le \frac{{{R_S}^n}}{{{\lambda _S}}}}\\
0&{{\rm{else}}}
\end{array}} \right..
\end{align}
\begin{proof}
\emph{As suggested in \cite{YetWCL}, when $E$ is randomly distributed in $S_p$, the CDF of $d_E$ can be given as}
\begin{align}
{F_{{d_E}}}\left( x \right) &= \int\limits_0^x {\int\limits_0^\pi  {\int\limits_0^{2\pi } {\frac{3}{{4\pi {R_S}^3}}d_E^2\sin {\phi _E}d{\theta _E}d{\phi _E}d\left( {{d_E}} \right)} } } \notag\\
&= \left\{ {\begin{array}{*{20}{l}}
0&{x < 0}\\
{\frac{{{x^3}}}{{R_S^3}}}&{0 \le x \le {R_S}}\\
1&{x > {R_S}}
\end{array}} \right..
\end{align}

\emph{So, one can have the CDF of $d_E^n/{\lambda _S}$ as}
\begin{align}
{F_{\frac{{d_E^n}}{{{\lambda _S}}}}}\left( x \right) &= \left\{ {\frac{{d_E^n}}{{{\lambda _S}}} \le x} \right\}\notag\\
& = \left\{ {{d_E} \le \sqrt[n]{{{\lambda _S}x}}} \right\}\notag\\
& = \left\{ {\begin{array}{*{20}{l}}
{0,}&{x < 0}\\
{\frac{{{\lambda _S}^{\frac{3}{n}}}}{{R_S^3}}{x^{\frac{3}{n}}}}&{0 \le x \le \frac{{{R_S}^n}}{{{\lambda _S}}}}\\
1&{x \ge \frac{{{R_S}^n}}{{{\lambda _S}}}}
\end{array}} \right..
\end{align}

\emph{Thus, one can easily obtain the PDF of $d_E^n/{\lambda _S}$ as it is the derivative of Eq. \eqref{pdfdEn2}.}

\emph{Then, the proof is completed.}
\end{proof}
\end{lemma}

Observing Eq. \eqref{SNRE}, and using Eqs. \eqref{pdfrayleigh} and \eqref{pdfdEn2}, the PDF of $\gamma_{E}$ can be calculated as
\begin{align}\label{pdfgammaE2}
{f_{{\gamma _E}}}\left( x \right) &= \int\limits_0^\infty  {y{f_{{{\left| {{h_{SE}}} \right|}^2}}}\left( {yx} \right)} {f_{\frac{{d_E^n}}{{{\lambda _S}}}}}\left( y \right)dy\notag\\
& = \frac{{3{\lambda _S}^{\frac{3}{n}}}}{{nR_S^3{g_{SE}}}}\int\limits_0^{\frac{{{R_S}^n}}{{{\lambda _S}}}} {{y^{\frac{3}{n}}}\exp \left( { - \frac{{yx}}{{{g_{SE}}}}} \right)} dy\notag\\
& = \frac{{3{\lambda _S}^{\frac{3}{n}}{g_{SE}}^{\frac{3}{n}}}}{{nR_S^3}}{x^{ - \frac{3}{n} - 1}}\gamma \left( {1 + \frac{3}{n},\frac{{{R_S}^n}}{{{\lambda _G}{g_{SE}}}}x} \right)\notag\\
& = {B_E}{x^{ - \frac{3}{n} - 1}}G_{1,2}^{1,1}\left[ {{C_E}x\left| {_{1 + \frac{3}{n},0}^1} \right.} \right],
\end{align}
where $B_E = \frac{{3({{\lambda _S}g_{SE}})^{\frac{3}{n}}}}{{nR_S^3}}$ and
${C_E} = \frac{{{R_S}^n}}{{{\lambda _S}{g_{SE}}}}$.

\begin{lemma}\label{lemma3}
{When $G$ is uniformly distributed in the circle with radius $R_C$, the PDF of $d_G^n/{\lambda _S}$ can presented as}
\begin{align}\label{pdfdgn}
{f_{\frac{{d_G^n}}{{{\lambda _S}}}}}\left( x \right) = \left\{ {\begin{array}{*{20}{l}}
{\frac{{2{\lambda _S}^{\frac{2}{n}}}}{{nR_C^2}}{x^{\frac{2}{n} - 1}}}&{{\rm{if~ }}\frac{{{h^n}}}{{{\lambda _S}}} \le x \le \frac{{{{\left( {{R_C}^2 + {h^2}} \right)}^{\frac{n}{2}}}}}{{{\lambda _S}}}}\\
0&{{\rm{else}}}
\end{array}} \right..
\end{align}
\begin{proof}
\emph{When $G$ is uniformly distributed in the circle with radius $R_C$, as shown in \cite{GPanCL2017}, one can have the PDF of $r$ as}
\begin{align}\label{fr}
{f_r}\left( x \right) = \frac{{2x}}{{R_C^2}}.
\end{align}

\emph{As $d_G^2 = {{h}^2}{\rm{ + }}{r^2}$ and using Eq. \eqref{fr}, the CDF of $d_G^{} = \sqrt {{{h}^2}{\rm{ + }}{r^2}} $ can be presented as}
\begin{align}
{F_{{d_G}}}\left( x \right) = \frac{{{x^2}}}{{R_C^2}} - \frac{{{{h}^2}}}{{R_C^2}}, ~{h} \le x \le \sqrt {{R_C}^2 + {{h}^2}}{={R_G}}.
\end{align}

\emph{Therefore, one can have the CDF of $d_G^n/{\lambda _S}$ as}
\begin{align}\label{CDFdgn}
{F_{\frac{{d_G^n}}{{{\lambda _S}}}}}\left( x \right) &= \left\{ {\frac{{d_G^n}}{{{\lambda _S}}} \le x} \right\}\notag\\
& = \left\{ {{d_G} \le \sqrt[n]{{{\lambda _S}x}}} \right\}\notag\\
& = \left\{ {\begin{array}{*{20}{l}}
0&{x \le \frac{{{h^n}}}{{{\lambda _S}}}}\\
{\frac{{{\lambda _S}^{\frac{2}{n}}}}{{R_C^2}}{x^{\frac{2}{n}}} - \frac{{{h^2}}}{{R_C^2}}}&{\frac{{{h^n}}}{{{\lambda _S}}} \le x \le \frac{{{{\left( {{R_C}^2 + {h^2}} \right)}^{\frac{n}{2}}}}}{{{\lambda _S}}}}\\
1&{x \ge \frac{{{{\left( {{R_C}^2 + {h^2}} \right)}^{\frac{n}{2}}}}}{{{\lambda _S}}}}
\end{array}} \right..
\end{align}

\emph{So, the PDF of $d_G^n/{\lambda _S}$ can be obtained as it is the derivative of Eq. \eqref{CDFdgn}.}

\emph{Then, the proof is completed.}
\end{proof}
\end{lemma}

{So, using \cite[Eq. (3.351.1)]{Gradshteyn}, and Eqs. \eqref{pdfrayleigh} and \eqref{pdfdgn}, the PDF of $\gamma_G$ can be calculated as}
\begin{align}
{f_{{\gamma _G}}}\left( x \right) &= \int\limits_0^\infty  {y{f_{{{\left| {{h_{SG}}} \right|}^2}}}\left( {yx} \right)} {f_{\frac{{d_G^n}}{{{\lambda _S}}}}}\left( y \right)dy\notag\\
 &= \frac{1}{{{g_{SG}}}}\int\limits_{\frac{{{h^n}}}{{{\lambda _S}}}}^{\frac{{{{\left( {{R_C}^2 + {h^2}} \right)}^{\frac{n}{2}}}}}{{{\lambda _S}}}} y \exp \left( { - \frac{{yx}}{{{g_{SG}}}}} \right)\frac{{2{\lambda _S}^{\frac{2}{n}}}}{{nR_C^2}}{y^{\frac{2}{n} - 1}}dy\notag\\
 &= \frac{{2{\lambda _S}^{\frac{2}{n}}}}{{nR_C^2{g_{SG}}}}\int\limits_0^{\frac{{{{\left( {{R_C}^2 + {h^2}} \right)}^{\frac{n}{2}}}}}{{{\lambda _S}}}} {{y^{\frac{2}{n}}}} \exp \left( { - \frac{{yx}}{{{g_{SG}}}}} \right)dy -\frac{{2{\lambda _S}^{\frac{2}{n}}}}{{nR_C^2{g_{SG}}}}\int\limits_0^{\frac{{{h^n}}}{{{\lambda _S}}}} {{y^{\frac{2}{n}}}} \exp \left( { - \frac{{yx}}{{{g_{SG}}}}} \right)dy\notag\\
 &= {C_G}{x^{ - \frac{2}{n} - 1}}G_{1,2}^{1,1}\left( {\left. {\frac{{{{\left( {{R_C}^2 + {h^2}} \right)}^{\frac{n}{2}}}}}{{{\lambda _S}{g_{SG}}}}x} \right|\begin{array}{*{20}{c}}
1\\
{1 + \frac{2}{n},0}
\end{array}} \right) -{C_G}{x^{ - \frac{2}{n} - 1}}G_{1,2}^{1,1}\left( {\left. {\frac{{{h^n}}}{{{\lambda _S}{g_{SG}}}}x} \right|\begin{array}{*{20}{c}}
1\\
{1 + \frac{2}{n},0}
\end{array}} \right),
\end{align}
where ${C_G} = \frac{{2({\lambda _S}{g_{SG}})^{\frac{2}{n}}}}{{nR_C^2}}$.

Accordingly, the CDF of $\gamma_G$ can be easily obtained as
\begin{align}\label{CDFgammaG}
{F_{{\gamma _G}}}\left( x \right) &= \int\limits_0^x {{f_{{\gamma _G}}}\left( y \right)} dy\notag\\
& = {C_G}\int\limits_0^x {{y^{ - \frac{2}{n} - 1}}G_{1,2}^{1,1}\left( {\left. {\frac{{{{\left( {{R_C}^2 + {h^2}} \right)}^{\frac{n}{2}}}}}{{{\lambda _S}{g_{SG}}}}y} \right|\begin{array}{*{20}{c}}
1\\
{1 + \frac{2}{n},0}
\end{array}} \right)} dy \notag\\
 &~~- {C_G}\int\limits_0^x {{y^{ - \frac{2}{n} - 1}}G_{1,2}^{1,1}\left( {\left. {\frac{{{h^n}}}{{{\lambda _S}{g_{SG}}}}y} \right|\begin{array}{*{20}{c}}
1\\
{1 + \frac{2}{n},0}
\end{array}} \right)} dy\notag\\
& = {C_G}{x^{ - \frac{2}{n}}}G_{2,3}^{1,2}\left( {\left. {\frac{{{{\left( {{R_C}^2 + {h^2}} \right)}^{\frac{n}{2}}}}}{{{\lambda _S}{g_{SG}}}}x} \right|\begin{array}{*{20}{c}}
{1 + \frac{2}{n},1}\\
{1 + \frac{2}{n},0,\frac{2}{n}}
\end{array}} \right) - {C_G}{x^{ - \frac{2}{n}}}G_{2,3}^{1,2}\left( {\left. {\frac{{{h^n}}}{{{\lambda _S}{g_{SG}}}}x} \right|\begin{array}{*{20}{c}}
{1 + \frac{2}{n},1}\\
{1 + \frac{2}{n},0,\frac{2}{n}}
\end{array}} \right) \notag\\
& ={C_G}{x^{ - \frac{2}{n}}}G_{2,3}^{1,2}\left[ {{A_G}x\left| {_{1 + \frac{2}{n},0,\frac{2}{n}}^{1 + \frac{2}{n},1}} \right.} \right]  - {C_G}{x^{ - \frac{2}{n}}}G_{2,3}^{1,2}\left[ {{B_G}x\left| {_{1 + \frac{2}{n},0,\frac{2}{n}}^{1 + \frac{2}{n},1}} \right.} \right],
\end{align}
where ${A_G} = \frac{{{{\left( {R_C^2 + {h^2}} \right)}^{\frac{n}{2}}}}}{{{\lambda _S}{g_{SG}}}}$ and
${B_G} = \frac{{{h^n}}}{{{\lambda _S}{g_{SG}}}}$.

Therefore, utilizing Eqs. \eqref{pdfgammaE2} and \eqref{CDFgammaG}, ${I_{Sp}} = \int\limits_{{V_{Sp}}} {{\rm{Pr\{ }}{\gamma _G} \le {\gamma _E}{\rm{\} }}} d{V_{{S_1}}}$ can be further derived as
\begin{align}\label{ISP}
{I_{SP}} &= \int\limits_{{V_{Sp}}} {{\rm{Pr}}\{ {\gamma _G} \le \Theta {\gamma _E}\} } d{V_{{S_1}}}\notag\\
& = \int\limits_0^\infty  {{F_{{\gamma _G}}}\left( {\Theta x} \right)} {f_{{\gamma _E}}}\left( x \right)dx\notag\\
& = {C_G}{\Theta ^{ - \frac{2}{n}}}\int\limits_0^\infty  {{x^{ - \frac{2}{n}}}G_{2,3}^{1,2}\left[ {{A_G}\Theta xx\left| {_{1 + \frac{2}{n},0,\frac{2}{n}}^{1 + \frac{2}{n},1}} \right.} \right]{f_{{\gamma _E}}}\left( x \right)dx}  - {C_G}\int\limits_0^\infty  {{x^{ - \frac{2}{n}}}G_{2,3}^{1,2}\left[ {{B_G}\Theta x\left| {_{1 + \frac{2}{n},0,\frac{2}{n}}^{1 + \frac{2}{n},1}} \right.} \right]{f_{{\gamma _E}}}\left( x \right)dx} \notag\\
& = {C_G}{B_E}{\Theta ^{\frac{3}{n}}}{A_G}^{\frac{5}{n}}G_{4,4}^{3,2}\left[ {\frac{{{C_E}}}{{{A_G}\Theta }}\left| {_{1 + \frac{3}{n},\frac{3}{n},\frac{5}{n},0}^{1,\frac{3}{n},1 + \frac{5}{n},1 + \frac{3}{n}}} \right.} \right] - {C_G}{B_E}{\Theta ^{\frac{3}{n}}}{B_G}^{\frac{5}{n}}G_{4,4}^{3,2}\left[ {\frac{{{C_E}}}{{{B_G}\Theta }}\left| {_{1 + \frac{3}{n},\frac{3}{n},\frac{5}{n},0}^{1,\frac{3}{n},1 + \frac{5}{n},1 + \frac{3}{n}}} \right.} \right]
\end{align}

\subsection{The Derivation of ${I_{{S_2}}}$}
\begin{lemma}\label{lemma4}
{As shown in Fig. \ref{fig_3}, when $E$ is uniformly distributed in the spherical cap with height ($R_S - h$), the PDF of $d_E^n/{\lambda _S}$ can presented as}
\begin{align}\label{pdfdEn3}
&{f_{\frac{{d_E^n}}{{{\lambda _S}}}}}\left( x \right) = \left\{ {\begin{array}{*{20}{l}}
{\frac{{2\pi }}{{n{V_S}}}\left( {{\lambda _S}^{\frac{3}{n}}{x^{\frac{3}{n} - 1}} - {h}{\lambda _S}^{\frac{2}{n}}{x^{\frac{2}{n} - 1}}} \right)}&{{\rm{if ~}}\frac{{h^n}}{{{\lambda _S}}} \le x \le \frac{{R_G^n}}{{{\lambda _S}}}}\\
0&{{\rm{else}}}
\end{array}} \right..
\end{align}
\begin{proof}
\emph{When $E$ is randomly distributed in $S_2$, the CDF of $d_E$ can be given as}
\begin{align}\label{CDFdEn}
{F_{{d_E}}}\left( x \right) &= \frac{1}{{{V_S}}}\int_0^{\arccos \left( {\frac{{{h}}}{x}} \right)} {\int_{\frac{{{h}}}{{\cos \left( \phi  \right)}}}^x {\int_0^{2\pi } {{\delta ^2}\sin \left( \phi  \right)d\theta d\delta d\phi } } } \notag\\
& = \frac{{2\pi }}{{{V_S}}}\int_0^{\arccos \left( {\frac{{{h}}}{x}} \right)} {\int_{\frac{{{h}}}{{\cos \left( \phi  \right)}}}^x {{\delta ^2}\sin \left( \phi  \right)d\delta d\phi } } \notag\\
& = \frac{{2\pi }}{{3{V_S}}}\int_0^{\arccos \left( {\frac{{{h}}}{x}} \right)} {\left( {{x^3} - h^3{{\sec }^3}\left( \phi  \right)} \right)\sin \left( \phi  \right)d\phi } \notag\\
& = \frac{\pi }{{3{V_S}}}\left( {2{x^3} - 3{h}{x^2} + h^3} \right).
\end{align}

\emph{So, the CDF of $d_E^n/{\lambda _S}$ can be obtained as}
\begin{align}\label{CDFdEn22}
{F_{\frac{{d_E^n}}{{{\lambda _S}}}}}\left( x \right) &= \Pr \left\{ {\frac{{d_E^n}}{{{\lambda _S}}} < x} \right\}\notag\\
& = \Pr \left\{ {{d_E} < {{\left( {{\lambda _S}x} \right)}^{\frac{1}{n}}}} \right\}\notag\\
& = \frac{\pi }{{3{V_S}}}\left( {2{{\left( {{{\left( {{\lambda _S}x} \right)}^{\frac{1}{n}}}} \right)}^3} - 3{h}{{\left( {{{\left( {{\lambda _S}x} \right)}^{\frac{1}{n}}}} \right)}^2} + h^3} \right)\notag\\
& = \frac{\pi }{{3{V_S}}}\left( {2{\lambda _S}^{\frac{3}{n}}{x^{\frac{3}{n}}} - 3{h}{\lambda _S}^{\frac{2}{n}}{x^{\frac{2}{n}}} + h^3} \right).
\end{align}

\emph{Thus, the the PDF of $d_E^n/{\lambda _S}$ can be derived via performing differential operation on Eq. \eqref{CDFdEn22}.}

\emph{Then, the proof is completed.}
\end{proof}
\end{lemma}

Observing Eq. \eqref{SNRE}, and using \cite[Eq. (3.351.1)]{Gradshteyn}, Eqs. \eqref{pdfrayleigh} and \eqref{pdfdEn3}, the PDF of $\gamma_{E}$ can be calculated as
\begin{align}\label{pdfgammaE2}
{f_{{\gamma _E}}}\left( x \right) &= \int\limits_0^\infty  {y{f_{{{\left| {{h_{SE}}} \right|}^2}}}\left( {yx} \right)} {f_{\frac{{d_E^n}}{{{\lambda _S}}}}}\left( y \right)dy\notag\\
& = {E_1}\int_{\frac{{h^n}}{{{\lambda _S}}}}^{\frac{{R_G^n}}{{{\lambda _S}}}} {{y^{\frac{3}{n}}}\exp \left( { - \frac{{yx}}{{{g_{SE}}}}} \right)dy} \notag \\
&~~~~ - {E_2}\int_{\frac{{h^n}}{{{\lambda _S}}}}^{\frac{{R_G^n}}{{{\lambda _S}}}} {{y^{\frac{2}{n}}}\exp \left( { - \frac{{yx}}{{{g_{SE}}}}} \right)dy} \notag\\
& = {E_1}{g_{SE}}^{\frac{3}{n} + 1}{f_1}\left( {\frac{3}{n}} \right) - {E_2}{g_{SE}}^{\frac{2}{n} + 1}{f_1}\left( {\frac{2}{n}} \right),
\end{align}

where ${E_1} = \frac{{2\pi {\lambda _S}^{\frac{3}{n}}}}{{n{V_S}{g_{SG}}}}$, ${E_2} = \frac{{2\pi {h}{\lambda _S}^{\frac{2}{n}}}}{{n{V_S}{g_{SG}}}}$ and ${f_1}\left( a \right) = \int_{\frac{{h^n}}{{{\lambda _S}}}}^{\frac{{R_S^n}}{{{\lambda _S}}}} {{y^a}\exp \left( { - \frac{{yx}}{{{g_{SG}}}}} \right)dy} $.

Therefore, using Eqs. \eqref{CDFgammaG} and \eqref{pdfgammaE2}, and the integral equation of Meijer's $G$-function presented in \cite{wolfram}, ${I_{{S_2}}} = \int\limits_{{V_{{S_2}}}} {{\rm{Pr\{ }}{\gamma _G} \le {\gamma _E}{\rm{\} }}} d{V_{{S_2}}}$ can be further derived as
\begin{align}\label{IS2}
{I_{{S_2}}} &= \int\limits_0^\infty  {{F_{{\gamma _G}}}\left( {\Theta x} \right)} {f_{{\gamma _E}}}\left( x \right)dx\notag\\
& = {C_G}{\Theta ^{ - \frac{2}{n}}}\int_0^\infty  {{x^{ - \frac{2}{n}}}G_{2,3}^{1,2}\left[ {{A_G}\Theta x\left| {_{1 + \frac{2}{n},0,\frac{2}{n}}^{1 + \frac{2}{n},1}} \right.} \right]{f_{{\gamma _E}}}\left( x \right)dx}\notag \\
& - {C_G}{\Theta ^{ - \frac{2}{n}}}\int_0^\infty  {{x^{ - \frac{2}{n}}}G_{2,3}^{1,2}\left[ {{A_G}\Theta x\left| {_{1 + \frac{2}{n},0,\frac{2}{n}}^{1 + \frac{2}{n},1}} \right.} \right]{f_{{\gamma _E}}}\left( x \right)dx} \notag\\
& = {C_G}{\Theta ^{ - \frac{2}{n}}}\left( {{f_2}\left( {{A_G}} \right) - {f_2}\left( {{B_G}} \right)} \right),
\end{align}

where
\begin{align}\label{H232}
{f_2}\left( t \right) &= \int_0^\infty  {{x^{ - \frac{2}{n}}}G_{2,3}^{1,2}\left[ {t\Theta x\left| {_{1 + \frac{2}{n},0,\frac{2}{n}}^{1 + \frac{2}{n},1}} \right.} \right]{f_{{\gamma _E}}}\left( x \right)dx} \notag\\
& = {E_1}{g_{SG}}^{\frac{3}{n} + 1}\int_0^\infty  {{f_1}\left( {\frac{3}{n}} \right){x^{ - \frac{2}{n}}}G_{2,3}^{1,2}\left[ {t\Theta x\left| {_{1 + \frac{2}{n},0,\frac{2}{n}}^{1 + \frac{2}{n},1}} \right.} \right]dx} \notag\\
& - {E_2}{g_{SG}}^{\frac{2}{n} + 1}\int_0^\infty  {{f_1}\left( {\frac{2}{n}} \right){x^{ - \frac{2}{n}}}G_{2,3}^{1,2}\left[ {t\Theta x\left| {_{1 + \frac{2}{n},0,\frac{2}{n}}^{1 + \frac{2}{n},1}} \right.} \right]dx} \notag\\
& = {E_1}{g_{SG}}^{\frac{3}{n} + 1}\left( {{f_3}\left( {\frac{3}{n},t,{C_E}} \right) - {f_3}\left( {\frac{3}{n},t,{D_E}} \right)} \right)\notag\\
& - {E_2}{g_{SG}}^{\frac{2}{n} + 1}\left( {{f_3}\left( {\frac{2}{n},t,{C_E}} \right) - {f_3}\left( {\frac{2}{n},t,{D_E}} \right)} \right)
\end{align}
and
\begin{align}\label{H232}
{f_3}\left( {s,t,b} \right) &= \int_0^\infty  {{x^{ - s - 1 - \frac{2}{n}}}G_{1,2}^{1,1}\left[ {bx\left| {_{s + 1,0}^1} \right.} \right]} G_{2,3}^{1,2}\left[ {t\Theta x\left| {_{1 + \frac{2}{n},0,\frac{2}{n}}^{1 + \frac{2}{n},1}} \right.} \right]dx \notag \\
& = {b^{s + \frac{2}{n}}}G_{4,4}^{2,3}\left[ {\frac{{t\Theta }}{b}\left| {_{1 + \frac{2}{n},s + \frac{2}{n},0,\frac{2}{n}}^{1 + \frac{2}{n},1,\frac{2}{n},1 + s + \frac{2}{n}}} \right.} \right].
\end{align}

Finally, the SOP over the downlink, ${P_{out,{\rm{dn}}}}$, can be obtained by inserting Eqs. \eqref{ISP} and \eqref{IS2} into Eq. \eqref{Pdown}.

\section{Numerical Results}
{\color{black}{In this section, Monte Carlo simulations are carried out to validate our proposed analytical expressions for the SOP over both downlink and uplink.}} The main adopted parameters are set as {\color{black}{$R_S = 20$ m, $h = 10$ m, $R_G = 15$ m, $R_{s}$ = 0.1 bits/s/Hz, $\lambda_S = 5$ dB, $\lambda_G = 1.25$ dB and $g_{GE} = g_{SE} = 1.1$}}. Moreover, the coverage distance of the source UAV is set from tens of meters to hundreds of meters to reflect the practical scenarios of UAVs in civil applications. During each case, we run $10^5$ trials for the Monte Carlo simulations and also consider $10^5$ times of the realizations of the considered systems. Furthermore, all following numerical results are given while the mean value of the power gain over the link between $S$ and ${G}$ ($g_{GS}$ or $g_{SG}$) increasing, {\color{black}{to show the system performance that can be achieved in each potential channel situations}}.

\subsection{Secrecy Outage over The Uplink}

In this subsection, we will investigate the secrecy outage performance over the uplink of the considered system shown in Fig. \ref{fig_2}.

\begin{figure}[!htb]
\centering
\includegraphics[width= 3.5in]{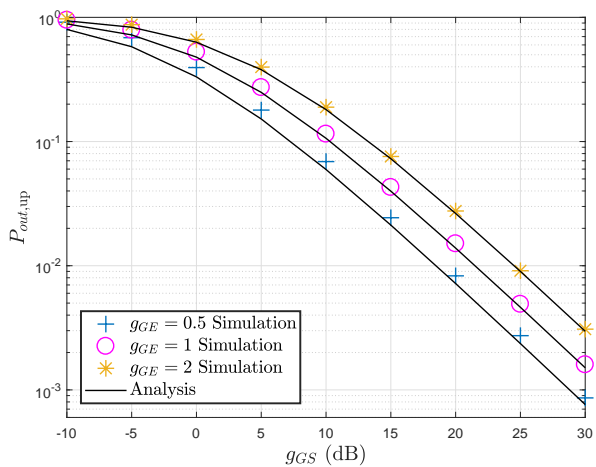}
\caption{Secrecy outage vs. ${g_{GS}}$ for various ${g_{GE}}$}
\label{fig_5}
\end{figure}

In Fig. \ref{fig_5}, numerical results are presented to address the influence of $g_{GE}$ on the secrecy outage performance of the considered system. We can see that the mean value of the power gain over $G-E$ link ($g_{GE}$) shows a negative effect on the secrecy outage, as a large $g_{GE}$ leads to degraded secrecy outage performance. Because a large $g_{GE}$ represents a high channel gain for the eavesdropping link from $G$ to $E$.

\begin{figure}[!htb]
\centering
\includegraphics[width= 3.5in]{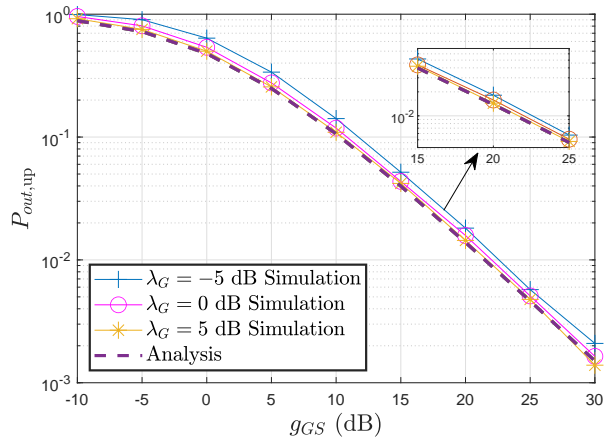}
\caption{Secrecy outage vs. ${g_{GS}}$ for various ${\lambda_S}$}
\label{fig_6}
\end{figure}

The effect of $\lambda_G$ (the transmit SNR at $G$) on the secrecy outage performance of the considered system is studied in Fig. \ref{fig_6}. $\lambda_G$ does not exhibit a significant influence on the secrecy outage performance, while a large $\lambda_G$ incurs improved secrecy outage performance. Moreover, as presented in Fig. \ref{fig_6}, the lines for the lower boundary of the SOP with  $\lambda_G = 0 $ and 5 dB are almost overlapped with each other. Then, we can have that increasing $\lambda_G$ is not an effective way to improve the secrecy outage performance of the considered system. {\color{black}{Because the changing of $\lambda_G$ will show the same trend and scale on the received SNR at $S$ and $E$, which can also be proved by Eqs. \eqref{SNRS} and \eqref{SNRE}.}}

\begin{figure}[!htb]
\centering
\includegraphics[width= 3.5in]{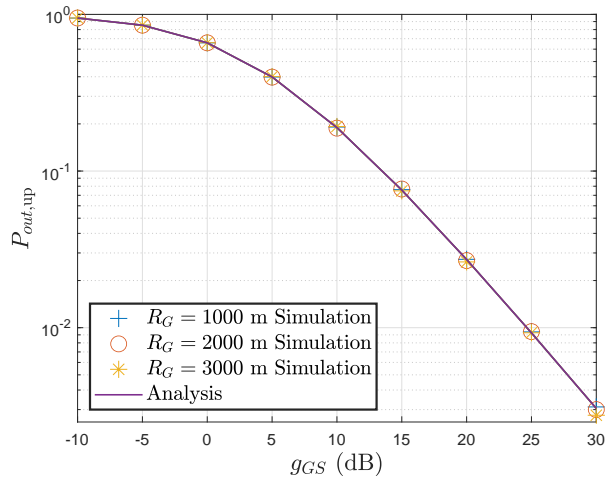}
\caption{Secrecy outage vs. ${g_{GS}}$ for various $R_G$}
\label{fig_7}
\end{figure}

Fig. \ref{fig_7} presents the secrecy outage performance of the considered system, while the radius of the coverage space of $G$, {\color{black}{$R_G= 1000$, 2000 and 3000 m}}\footnote{{\color{black}{In this work, the effects of the radius of the coverage space of $S$ and $G$ on the secrecy outage performance are considered. Because the radius of the coverage space of a transmitter not only depends on the transmit power at the transmitter but also is determined by the sensitivity of the receiver.}}}. One can easily observe that the $P_{out,{\rm{up}}}$ with various $R_G$ is similar. Moreover, we can also find that $R_G$ (the radius of the coverage space of the ground receiver, $G$) shows a negative effect on the secrecy outage performance, which is more apparent when $R_G$ is small. This comes from the fact that a large $R_G$ leads to a large coverage space of $G$, and then the probability that the distance $d_{GE}$ gets large increases, which finally results in bad channel quality of the eavesdropping link.

\begin{figure}[!htb]
\centering
\includegraphics[width= 3.5in]{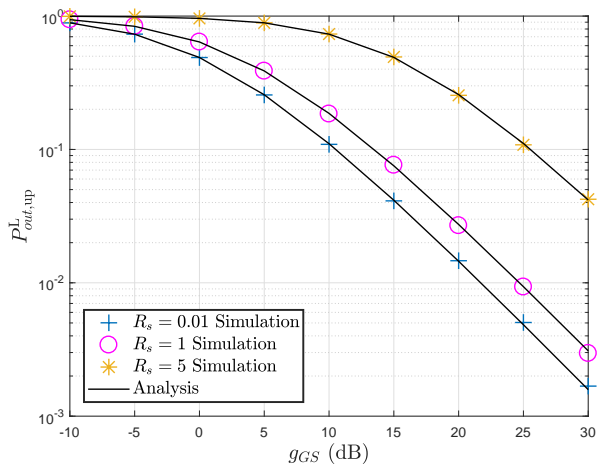}
\caption{Secrecy outage vs. ${g_{GS}}$ for various $R_s$}
\label{fig_8}
\end{figure}

In Fig. \ref{fig_8}, the secrecy outage performance of the considered system with various outage threshold ($R_s$) is studied. A large $R_s$ shows a large $P_{out,{\rm{up}}}$, which means the secrecy outage performance of the considered system gets worse. It is because that a large $R_s$ represents a large outage threshold.

Furthermore, it can be seen from Figs. \ref{fig_5}-\ref{fig_8} that increasing $g_{GS}$ can improve the secrecy outage performance of the considered system. Because a large $g_{GS}$ represents a high channel gain for the main link from $G$ to $S$. Also, simulation and analysis results match very well with each other, which verifies the correctness of our proposed analytical model.

\begin{figure}[!htb]
\centering
\includegraphics[width= 3.5in]{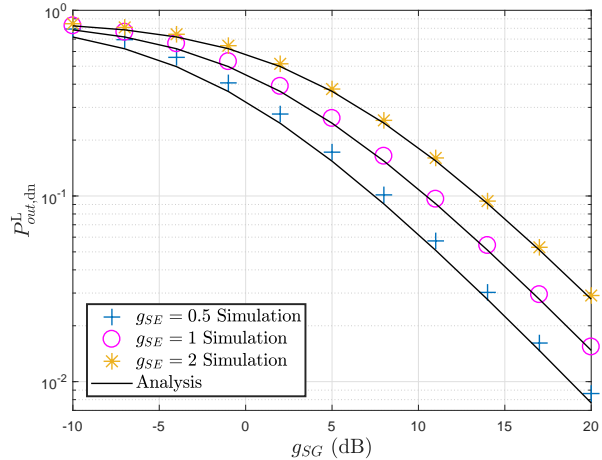}
\caption{Secrecy outage vs. ${g_{SG}}$ for various ${g_{GE}}$}
\label{fig_9}
\end{figure}

\begin{figure}[!htb]
\centering
\includegraphics[width= 3.5in]{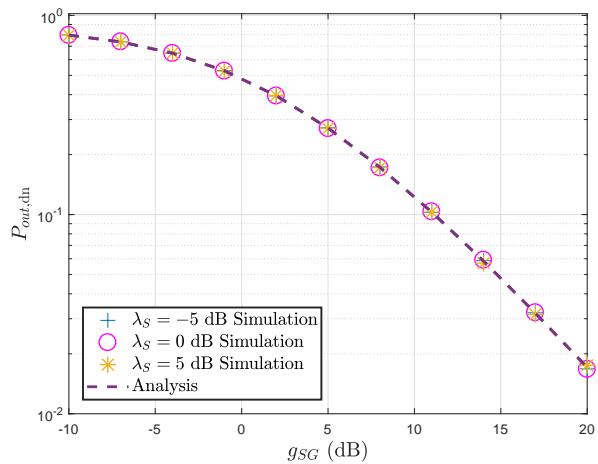}
\caption{Secrecy outage vs. ${g_{SG}}$ for various ${\lambda_S}$}
\label{fig_10}
\end{figure}

\subsection{Secrecy Outage over The Downlink}

In this subsection, the secrecy outage performance over the uplink of the considered system (shown in Fig. \ref{fig_3}) will be studied.

Fig. \ref{fig_9} depicts the numerical results of the secrecy outage performance with various the mean value of the power gain over $S-E$ link, $g_{SE}$, while $g_{SG}$ increasing. One can observe that the secrecy outage performance with a small $g_{SE}$ outperforms that with a large $g_{SE}$, as a large $g_{SE}$ means better channel gain for the eavesdropping link, which leads to more information being overheard.

In Fig. \ref{fig_10}, we investigate how the secrecy outage performance is influenced by $\lambda_S$ (the transmit SNR at $S$). The secrecy outage lines for different $\lambda_S$ (namely, $-5$, 0, and 5 dB) overlap with each other, which indicates that adjusting $\lambda_S$ cannot improve the secrecy outage performance of the considered system. This observation is similar to the one obtained for the uplink from Fig. \ref{fig_6}. Also, it can be explained that increasing $\lambda_S$ will improve the received SNR at both $G$ and $E$, and then the secrecy outage performance of the considered system cannot be improved anymore.

\begin{figure}[!htb]
\centering
\includegraphics[width= 3.5in]{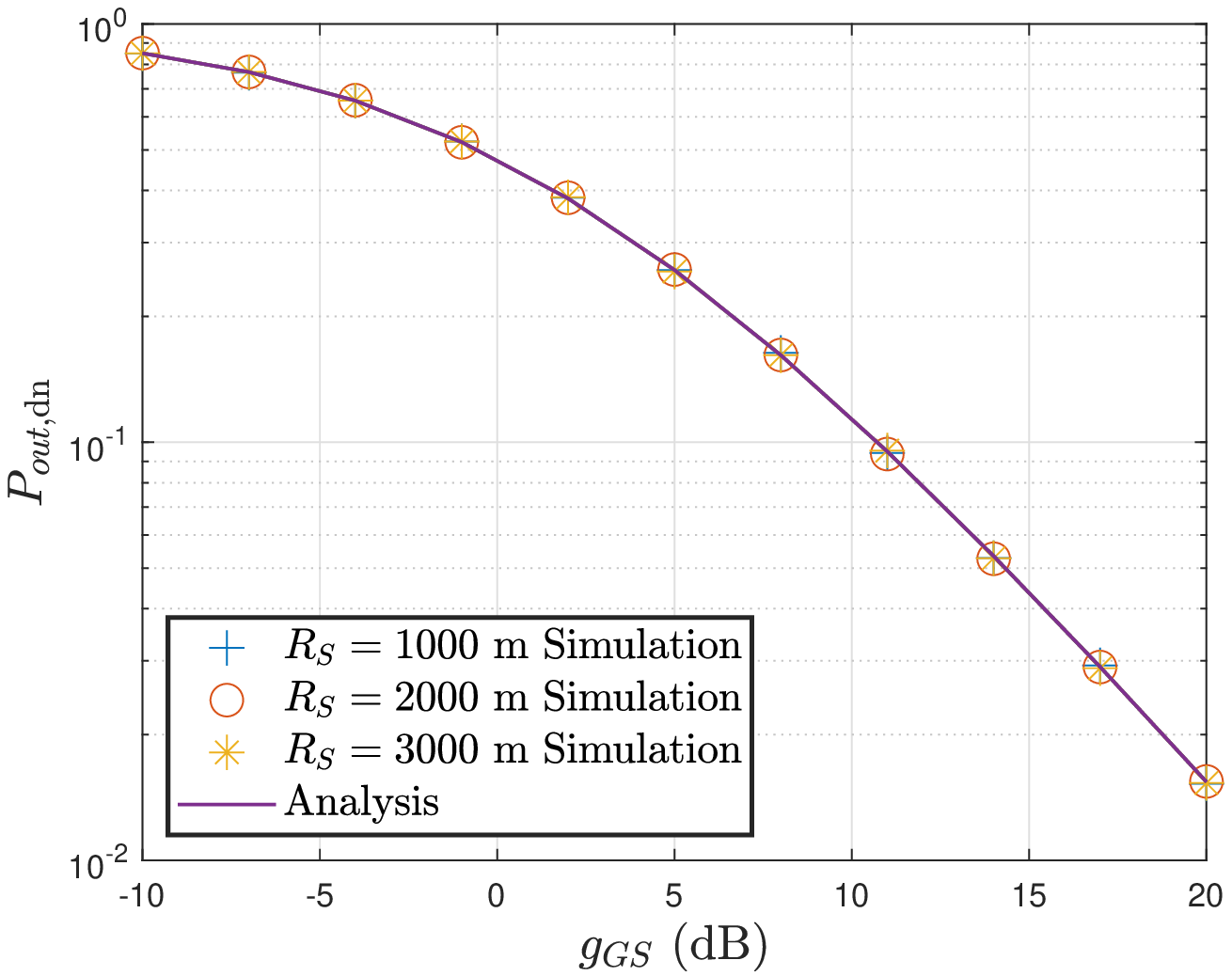}
\caption{Secrecy outage vs. ${g_{SG}}$ for various $R_S$}
\label{fig_11}
\end{figure}

The impact of $R_S$ (the radius of the coverage space of $S$) on the secrecy outage performance of the considered system is shown in Fig. \ref{fig_11}, while $g_{SG}$ increasing. It is shown that the curves of $P_{out,{\rm{dn}}}$ for {\color{black}{$R_S = 1000$, 2000 and 3000 m}} fully overlap with each other, which reveals that $R_S$ has no influence on the secrecy outage performance of the considered system, {\color{black}{because adjusting $R_s$ exhibits a same influence on the transmission distances of $S-E$ and $S-G$ links.}} This finding is similar to the one achieved from Fig. \ref{fig_7}, both of which demonstrate that enlarging the coverage space of the transmitter cannot improve the secrecy outage performance.

\begin{figure}[!htb]
\centering
\includegraphics[width= 3.5in]{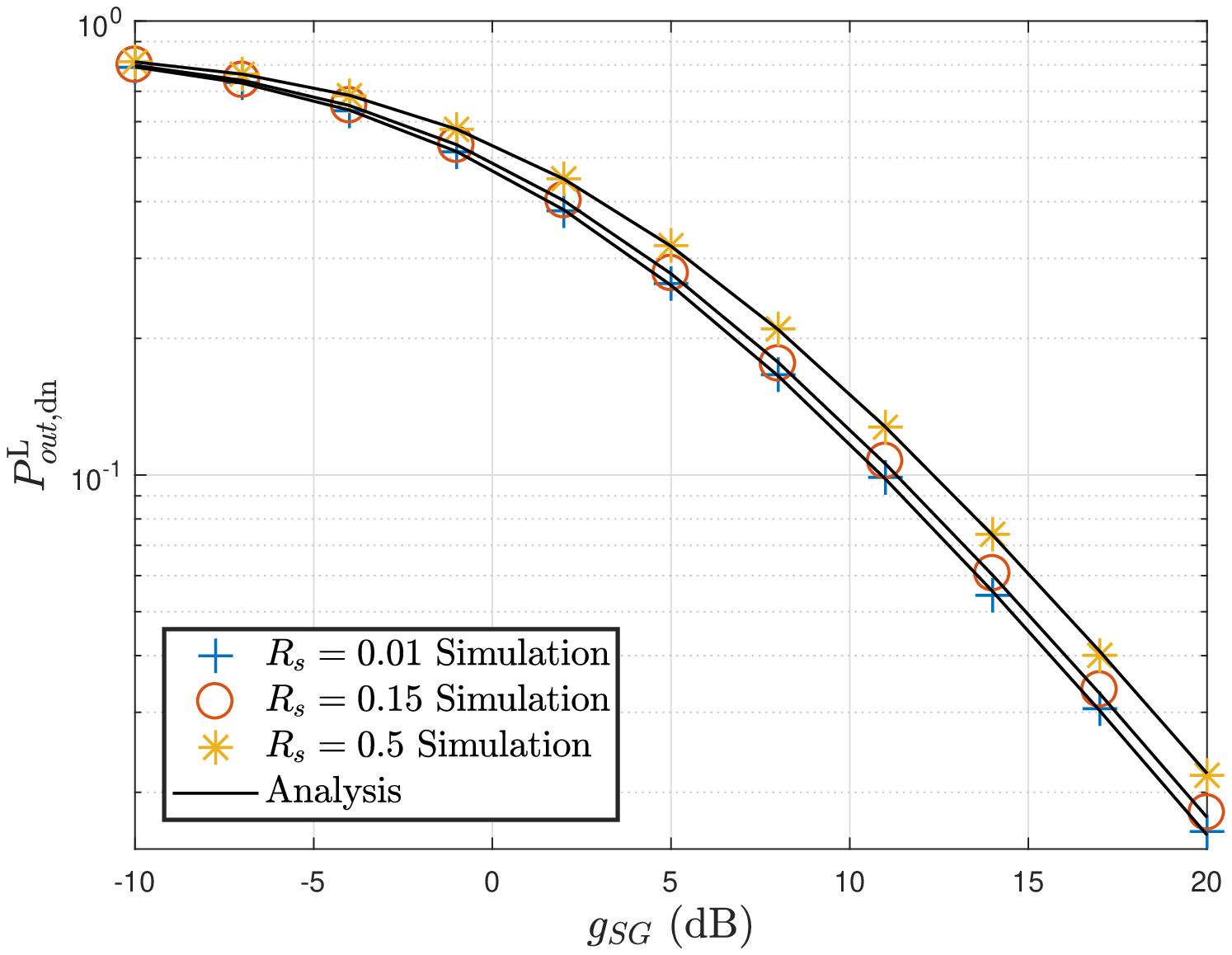}
\caption{Secrecy outage vs. ${g_{SG}}$ for various $R_s$}
\label{fig_12}
\end{figure}
\begin{figure}[!htb]
\centering
\includegraphics[width= 3.5in]{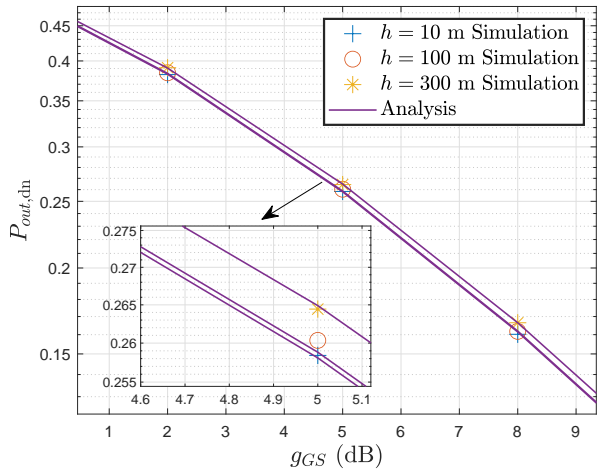}
\caption{{\color{black}{Secrecy outage vs. ${g_{SG}}$ for various $h$}}}
\label{fig_13}
\end{figure}

Fig. \ref{fig_12} presents the secrecy outage performance of the considered system with various outage threshold ($R_s$). The results show that $P_{out,{\rm{dn}}}$ with a small $R_s$ outperforms that with a large $R_s$. Because a large $R_s$ implies that the probability that secrecy capacity of the considered system is less than the threshold will increases. This observation is as same as the one seen from Fig. \ref{fig_8} for the uplink.

{\color{black}{In Fig. \ref{fig_13}, the SOP lines for various the height of $S$, $h$, are depicted while $R_S = 1000$ m. Clearly, the height of $S$, $h$, shows a negative effect on SOP, because a large $h$ leads to a low received SNR at the ground receiver, $G$, and further incurs the degraded secrecy outage performance.}} {\color{black}{Though the secrecy performance will degrade while the height of the UAV increases, optimizations can be set up to realize optimal secrecy performance vis choosing suitable heights for the UAV, and some schemes can be designed to safeguard the security of UAV system, e.g., jamming and artificial noise schemes.}}

Finally, as presented in Figs. \ref{fig_9}-\ref{fig_13}, it is noted that $g_{SG}$ exhibits a similar effect on the secrecy outage performance of the considered system as $g_{GS}$ does in Figs. \ref{fig_5}-\ref{fig_8}. This can also be explained by the idea of the reason given at the end of the last subsection. Moreover, simulation and analysis results show perfect matching with each other, which demonstrates the accuracy of the proposed analysis model.

\section{Conclusion}
{\color{black}{In this paper, we have studied the secrecy outage performance of a UAV system with linear trajectory}} and derived the closed-form analytical expressions for the lower boundary of the SOP of both downlink and uplink. We consider the randomness of the positions of all UAVs and the ground receiver to make the considered system more practical.

Observing from the numerical results, we can reach some remarks as follows:

1) The radius of the coverage space of $S$ does not exhibit an apparent impact on the secrecy outage performance over the downlink, as well as that of $G$ showing a weak influence on the secrecy outage performance over the uplink.

2) The transmit SNR at $S$ and $G$ provides a very weak impact on the secrecy outage performance over the downlink and the uplink, respectively.

3) The height of $S$ exhibits a negative effect on the secrecy outage performance of the considered system.

In this work, we mainly deal with modeling and analyzing the secrecy outage performance of the target system. In future work, we will work on the methods to improve the secrecy outage performance of the target system. e.g., cooperative jamming and artificial noise.

\section*{Appendix: Proof of Theorem 1}
As $C$ is the midpoint of $AB$, it is easy to obtain $AC = l/2$. Then, we can have $\cos A = \frac{l}{{2b}}$. According to Cosine theorem, we can also achieve
\begin{align}
\cos A = \frac{{{b^2} + {x^2} - {y^2}}}{{2bx}}.
\end{align}

Solving this quadric equation leads
\begin{align}
x &= \frac{{l \pm \sqrt {4{y^2} + {l^2} - 4{b^2}} }}{2}\notag\\
& = g\left( y \right).
\end{align}

Since $S$ is uniformly distributed over $AB$, the PDF of $x$ is given as
\begin{align}
{f_x}\left( x \right) = \frac{1}{l}, 0 \le x \le l.
\end{align}

Using transformation of random variable, we can obtain the PDF of $y$ as
\begin{align}
{f_y}\left( y \right) &= {f_x}\left( {g\left( y \right)} \right)\left| {g'\left( y \right)} \right|\notag\\
 &= \left\{ {\begin{array}{*{20}{l}}
{\frac{{2y}}{{l\sqrt {4{y^2} + {l^2} - 4{b^2}} }}}&{{\rm{if}}\sqrt {b^2 - \frac{{{l^2}}}{4}}  \le y \le b}\\
0&{{\rm{else}}}
\end{array}} \right..
\end{align}

Then, the proof is completed.

\end{document}